\documentclass[pra,aps,twocolumn,nofootinbib,floatfix]{revtex4-1}
\usepackage{graphicx}
\usepackage{bm}
\usepackage{amssymb}

\def\be{\begin{equation}}
\def\ee{\end{equation}}
\def\ba{\begin{eqnarray}}
\def\ea{\end{eqnarray}}
\def\half{\frac{1}{2}}
\def\sgn{\mbox{sgn}}
\def\D{\Delta}
\def\tF{\tilde{F}}
\def\cF{{\cal F}}
\def\cL{{\cal L}}
\def\fren{\mathbb{F}}
\def\cur{\mathbb{I}}
\def\ome{\omega}
\def\bxi{\bar{\xi}}

\def\Im{\mbox{Im}}
\def\Ibif{I_{\rm bif}}
\def\qbif{q_{\rm bif}}
\def\Fact{\Delta F}
\begin{document}
\title{Metastability and bifurcation in superconducting nanorings}
\author{Sergei Khlebnikov}
\email{skhleb@purdue.edu}
\affiliation{Department of Physics and Astronomy, Purdue University, 
West Lafayette, IN 47907, USA}
\begin{abstract}
We describe an approach, based on direct numerical solution of the Usadel
equation, to finding stationary points of the free energy of superconducting nanorings.
We consider both uniform (equilibrium) solutions and the critical droplets that
mediate activated transitions between them. For the uniform solutions, we 
compute the critical current as a function of the temperature,
thus obtaining a correction factor to Bardeen's 1962 interpolation formula. For
the droplets, 
we present a metastability chart that shows the activation energy as a function of
the temperature and current. A comparison of the activation energy for a ring to 
experimental results for a wire connected to superconducting leads reveals a discrepancy
at large currents. We discuss possible reasons for it.
We also discuss the nature of the bifurcation point at which
the droplet merges with the uniform solution.
\end{abstract}
\maketitle
\setcounter{footnote}{1}
\section{Introduction and summary}

Destruction of superconductivity in thin wires at high currents is a classic topic.
Bardeen's 1962 review \cite{Bardeen} summarizes the state of the art at the time. 
In particular, it presents an interpolation formula
\be
I_c(T) \approx I_c(0) [1 - (T/T_c)^2]^{3/2} 
\label{bardeen}
\ee
for the critical (depairing) current as a function of the temperature.
An important development subsequent to Bardeen's article
has been Little's work \cite{Little}, which emphasized the role of 
large thermal fluctuations (phase slips) as a cause for 
transition to the normal state at currents below $I_c$.
Experimental studies of this switching transition have developed fast 
in recent years \cite{Sahu&al,Li&al,Aref&al}.

On the theoretical side, a study of thermal phase slips begins with
identifying the saddle-point 
of the free energy (the critical droplet) that determines the activation barrier.
A number of models
have been used for this purpose. The original computation \cite{LA}
of Langer and Ambegaokar (LA)
was in the context of the Ginzburg-Landau (GL) theory. The GL theory is well motivated
microscopically but is limited to a vicinity of the critical temperature. One alternative 
may be to use a discrete model---essentially, a chain of Josephson junctions. Relevance of
such models to nanowires has been discussed in 
\cite{Matveev&al,Goswami&Chakravarty,Kh:2016}.

Here, we proceed with a continuum description and
present results for activation energies obtained by
a direct numerical solution of the one-dimensional Usadel equation \cite{Usadel}. 
The latter arises
as the dirty-limit reduction of the Eilenberger-Larkin-Ovchinnikov theory 
\cite{Eilenberger,LO}. It
is well motivated microscopically, applies at any temperature, and contains the GL theory
as a limit. For an infinite (very long) wire, the critical droplets have been considered
on the basis of the Usadel equation in \cite{Semenov&al}. Here, we will be interested in
solutions for a wire of a finite length.

Usadel's equation is second-order in spatial derivatives and requires boundary
conditions. Let us say a few words about those.
Bardeen's formula (\ref{bardeen}) pertains to a uniform superconducting state:
the current density and the gap are the same everywhere along the wire 
(which is assumed here to be in the $x$ direction). If the wire is
connected by leads into an external circuit, such a uniform state can only be an
idealization, the more so the shorter the wire is. The true equilibrium in this case
is necessarily $x$-dependent. Such $x$-dependent solutions have been found for
very short wires (bridges of length $L \ll \xi$, where $\xi$ is the
coherence length), both in the GL theory \cite{AL} and on the basis of the 
Usadel equation \cite{KO}. In principle, the numerical approach we describe here
can be used to find such solutions also for longer wires, provided one is willing
to do some modeling of physics in the leads.\footnote{In the context of the GL
theory, transition to longer wires has been recently considered in 
\cite{Marychev&Vodolazov}.}

Another aspect of the boundary conditions, which is particularly relevant to a study
of phase slips, is whether the leads prevent rapid changes in the boundary values of the 
phase of the order parameter 
(as bulk superconductors do) or allow such changes to occur easily
(the case, for instance, for normal contacts). 
In the first case, the critical droplet does not have to have the same value of the
current as the equilibrium state from which it originates, while in the
second case it typically does.\footnote{An intuitive picture of how bulk 
superconducting leads allow for fluctuations of the current in the wire can be 
obtained by viewing the leads
as small impedances connecting the ends of the wire to the ground. An
impedance will be of order $Z= (\cL/ C)^{1/2}$, where $\cL$ is the kinetic inductance
(which is small for a large superconductor), and $C$ is the capacitance of 
(a portion of) the lead relative to the ground. 
The two impedances shunt the ends of the wire, allowing the current in
it to fluctuate.} 
For long wires, the two scenarios are not that different (as shown by McCumber
\cite{McCumber} in the
case of the LA saddle point), but for shorter wires
there is a genuine difference, 
reflecting the difference in the experimental setup. We refer to the first scenario
as nucleation at a fixed winding number, and to the second as nucleation at a fixed
current.

The simplest system in which nucleation of the droplet occurs at a fixed winding number,
rather than a fixed current, is a superconducting ring. One facet of that simplicity is
that, for a ring of a uniform cross section, there is a uniform equilibrium state regardless
of the length. This fortuitously circumvents the problem of finding the $x$-dependent 
equilibrium characteristic of short wires in the presence of leads. In particular,
results for a ring of a finite length can be directly compared to the predictions of 
Bardeen's formula (\ref{bardeen}), and a correction factor to (\ref{bardeen})
can be obtained. With this in mind,
we have chosen the ring geometry for the present study. To treat the winding number
as a continuous variable, we consider twisted boundary conditions,
such as would arise in the presence of a fractional magnetic flux through the ring.

\begin{figure}
\begin{center}
\includegraphics[width=3.4in]{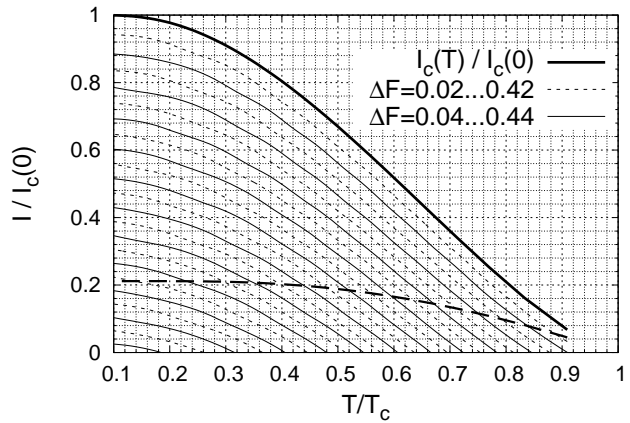}
\end{center}                                              
\caption{Metastability chart, 
showing level contours of the activation free energy $\Fact$
for a critical droplet
nucleating from a uniform equilibrium, as a function of the equilibrium 
current $I$ and the temperature. $\Fact$ is in units of $\fren_0$,
eq. (\ref{F_0}). The thick uppermost curve is the critical current, defined
as the maximum current possible for a uniform solution at a given temperature.
The levels of $\Fact$ increase
down and correspond to even (solid lines) or odd (dashed lines)
multiples of 0.02 (the innermost curve being for $\Fact = 0.44$). 
The thick dashed line corresponds to the ring biased by half the flux quantum.
Below that line, the uniform state is absolutely stable, and the droplet 
represents a fluctuation towards a uniform state with a higher free energy. 
The results are for a ring of half-length $L/2 = 41$ 
in units of the diffusion length (\ref{bxi}).
The effective coupling is $\lambda N(0) = 0.234$.
}                                              
\label{fig:chart}                                                                       
\end{figure}

After some preliminaries in Sec.~\ref{sec:prelim}, we discuss details of the boundary
conditions in Sec.~\ref{sec:bc} and then describe the numerical results in
Sec.~\ref{sec:num}. Here, we present in advance the main result: the metastability
chart (Fig.~\ref{fig:chart}), which shows level contours of the activation free 
energy
\be
\Fact = F({\rm droplet}) - F({\rm uniform})
\label{Fact}
\ee
for the critical droplet mediating transition from the uniform equilibrium 
with given values of the temperature and current to the state with one fewer unit of winding.
When $\Fact$ is expressed as a multiple of the free energy unit (\ref{F_0}), 
the metastability chart depends on only two parameters: the length $L$
of the ring and
the effective electron-electron coupling $\lambda N(0)$. For the typical weak-coupling 
case $\lambda N(0) < 0.3$, the dependence of $\Fact$ on the coupling is
rather weak. 

For a uniform ring, the properties of the equilibrium state do not depend on $L$,
but those of the critical droplet do. 
One might expect that, as $L$ becomes large compared to a suitably defined
coherence length
$\xi(T)$, $\Fact$ approaches the constant value
corresponding to an infinite wire, with finite-size corrections of 
order $\xi(T) / L$. That is true over most of the chart but not in a narrow
band of currents near the critical. The reason is that the size of the 
droplet grows for currents near $I_c(T)$,
so even for a small $\xi(T) / L$ 
there is a range of currents for which 
finite-size effects are important.\footnote{For the LA droplet in the GL theory, this
effect can be deduced already from the expressions presented in \cite{LA}.}
In what follows, we will often quote the length in units
of the diffusion length $\bxi$, defined by eq.~(\ref{bxi}) below. 
To estimate the significance of finite-size effects on $\Fact$, we can compare the results 
for $L = 82$, which is the value
used for the chart of Fig.~\ref{fig:chart}, to those for about twice
the length, $L=162$. We concentrate on temperatures $0.25 < T/T_c < 0.5$
and currents $I/I_c(T) \sim 0.9$, where thermal
switching transitions are typically observed. More specifically, we consider the 
``observability line'' $\Fact = 0.03$ (which lies in the middle of the second 
metastability band in Fig.~\ref{fig:chart}). We have found that, for these ranges of
the parameters, increasing the length to $L=162$ leads to a decrease in
$\Fact$ by about 20\%. 

We should remark that, at a finite $L$, the critical (maximum) 
current $I_c(T)$, represented by the thick upper 
curve in Fig.~\ref{fig:chart}, does not coincide with the bifurcation current
$\Ibif(T)$, at which the droplet merges with the uniform solution and the activation
barrier disappears. The two would be strictly the same only in the $L\to \infty$ limit.
The separation between $\Ibif(T)$ and $I_c(T)$ is possible because
the value of the current does 
not define the uniform solution uniquely: the winding number does. As we increase the
winding from zero, the current first increases until it reaches the maximum, 
$I_c(T)$, where
$\Fact$ is still nonzero, and then decreases down to $\Ibif(T)$, where $\Fact$ finally
vanishes.\footnote{Very recently, this effect has been observed experimentally
\cite{Petkovic&al}.} The value of $L$ chosen
for Fig.~\ref{fig:chart}, however, is already large enough for the difference 
between $I_c$ and $\Ibif$ to be unnoticeable on the scale of the plot. 
We return to discussion of this point in Sec.~\ref{sec:num},
where results for a smaller length are presented.

In Sec.~\ref{subsec:exp}, we carry out a comparison of results for $\D F$
obtained on the basis of our numerical solutions to the experimental results for
one of the samples of Ref.~\cite{Aref&al}. The comparison uses only experimentally
measured quantities and has no free parameters. We find a very good agreement 
at low currents, but a significant discrepancy at large ones, where switching 
transitions are observed. The experimentally determined $\D F$ scales as 
$(1- I/I_c)^{3/2}$ near the critical current, while the one obtained numerically is
almost linear. As a consequence, the numerical result there is significantly larger
than $\D F$ deduced from experiment. 
We mention a couple of possible reasons for this discrepancy in
Sec.~\ref{subsec:exp}. Here, we remark only that the resolution of the problem would
be much aided by experiments on rings thin enough for the ``premature'' switching 
($I < I_c$, $\D F \neq 0$) to have a chance to be observed.
The experiments \cite{Petkovic&al} on thicker rings have already accessed the
``deterministic'' switching regime $\D F = 0$.

For a uniform ring, the equilibrium solution continues to exist even after
the winding number $W$ is increased beyond the bifurcation point $W_{\rm bif}$, 
although it becomes absolutely unstable. One may ask if this property is generic
and will hold also for a wire connected 
to leads. In general, we consider that unlikely: 
as we discuss in Sec.~\ref{sec:top}, the persistence of the 
equilibrium solution for a ring can be seen as a consequence of 
the translational invariance. (More generally, is can be seen as a consequence of
there being different ``versions'' of the critical droplet that are degenerate in 
free energy.) It would nevertheless be interesting to see if there are any distinct 
physical consequences of such absolutely unstable solutions: on the one hand, fabricating 
a suitable ring sample may not be out of question; on the other, for a wire connected 
to leads, there is an approximate translation symmetry in the middle, so the same 
consequences may show up in that case as well.

\section{Preliminaries} \label{sec:prelim}
Usadel's equation \cite{Usadel}
can be obtained by variation of a certain free energy functional, 
which is essentially that
of an $O(3)$ nonlinear sigma model. The corresponding free energy density is 
\begin{widetext}
\be
\cF = \frac{|\D|^2}{\lambda} + 2\pi T N(0) \sum_{\omega > 0}
\left\{ - \D^* F - \D F^* + \half D \left[ \nabla F^* \nabla F + (\nabla G)^2 \right] 
+ 2\omega (1 - G) \right\} \, .
\label{fren}
\ee
\end{widetext}
The sigma-model variables are the complex $F(x,\ome)\equiv F$ and the real
$G(x,\ome) \equiv G$, related by the nonlinear constraint $|F|^2 + G^2 = 1$.
Variation with respect to $\D$ produces the self-consistency 
condition
\be
\D(x) = 2\pi T \lambda N(0) \sum_{\omega > 0} F(x, \ome) \, .
\label{scons}
\ee
In a thin superconductor of uniform cross section, $F$ and $G$ depend on only one 
spatial coordinate, $x$ (so in particular $\nabla = \partial_x$). 
In addition, they depend on the Matsubara frequency
$\omega$ that runs over the values $\omega_k = 2\pi T (k + \half)$, where $k$ is 
an integer. Note that the summations over $\omega$ in (\ref{fren}) and (\ref{scons}) 
are restricted to $k \geq 0$. If one
wishes to pursue the sigma-model context,
one can think of $F$ and $G$ as the in-plane and out-of-plane
components of a (fictitious) magnetization,
and of the superconducting gap $\D$ as the in-plane component of the self-consistent
fictitious magnetic field.

The parameters appearing in (\ref{fren}) are as follows: $\lambda$ is the 
electron-electron 
coupling, $T$ is the temperature, 
$N(0)$ is the density of states at the Fermi level (for a single spin projection), 
and $D$ is the electron diffusion constant. Following \cite{Eilenberger},
we will use the substitution
\be
\frac{1}{\lambda N(0)} = \ln \frac{T}{T_c} + \sum_{k = 0}^{K - 1} \frac{1}{k + \half}
\label{eilen}
\ee
to relate $\lambda N(0)$ to the critical temperature. The sum
in (\ref{eilen}) is cut off just below some integer $K$, which corresponds to the
maximum (Debye) frequency in units of $2\pi T$.

Variation of the free energy with respect to $F$ 
(with $G$ expressed as $G= (1 - |F|^2)^{1/2}$) reproduces the Usadel equation
\be
- \D - \frac{D}{2} \left(\nabla^2 F - \frac{F}{G} \nabla^2 G \right) 
+ \frac{\omega F}{G} = 0 \, ,
\label{usad}
\ee
which should be solved together with (\ref{scons}).

The current density, in units of twice the electron charge, is \cite{Usadel}
\be
J = - i \pi T N(0) D \sum_{\ome > 0} (F^* \nabla F - F \nabla F^*) \, .
\label{cur}
\ee
Although this nominally depends on $x$, for static solutions (the only case considered
here) in samples of uniform cross section, 
charge conservation ensures that $J$ is $x$-independent.

It is convenient to define a new field, $\tF(x,\omega)$, by removing ``most of the
winding'' from $F$ (we will say precisely how much below), as follows:
\be
F(x,\ome) = e^{iqx} \tF(x,\ome) =  e^{iqx} [R(x,\ome) + i I (x,\ome)] \, ,
\label{tF}
\ee
where, for now, $q$ remains unspecified. 
For future use, we have separated the real and imaginary parts of $\tF$.
In terms of $\tF$, the gradient term in (\ref{fren})
becomes
\be
\nabla F^* \nabla F = (\nabla - i q) \tF^* (\nabla + i q) \tF \, ,
\label{grad_tF}
\ee
and the nonlinear constraint becomes
\be
|\tF|^2 + G^2 = 1 \, .
\label{nonlin}
\ee
In what follows, we consider $\tF$, rather than $F$, as the main
dynamical variable and $q$ as a parameter, on which the free energy depends 
explicitly via (\ref{grad_tF}).

\section{Boundary conditions} \label{sec:bc}
We begin by
considering the theory on a ring of length $L$, with periodic conditions on 
$F(x,\omega)$ for all $\omega$. Later, we will allow for twisted boundary conditions,
such as those associated with a magnetic flux through the ring.

For our purposes, it is not sufficient to simply state that the solution is periodic.
For one thing,
because of the perfect translational invariance, any $x$-dependent solution will
have a translational
zero mode (a zero eigenvalue of the Hessian matrix), which will render the
numerical method we use inapplicable. The same applies to the zero mode generated by 
the symmetry with respect to global phase rotations. We wish to
eliminate these zero modes by placing the ``core'' of the solution at a particular
point (say, $x=0$) and also fixing its overall phase.

We first note that the free energy is invariant under the discrete transformation 
\be
P: F(x,\omega) \to F^*(-x, \omega) \, .
\label{P}
\ee
To eliminate the zero modes, we concentrate on solutions 
that have definite parity under $P$. It is then sufficient to consider only 
$P$-odd solutions, namely those that satisfy
\be
F(x,\omega) = - F^*(-x,\omega) \, ,
\label{odd}
\ee
because any $P$-even solution can be turned into a $P$-odd one by multiplication
by $i$. The condition (\ref{odd}) breaks the translational symmetry to translation
by $L/2$, and the symmetry with respect to global phase rotations to multiplication
by $-1$. These are discrete symmetries, which do not produce zero modes.

As an example of how (\ref{odd}) fixes the overall phase, consider one of the
uniform solutions, which are of the form
\be
F(x,\omega) = f(\omega) e^{iQ x} \, ,
\label{uni}
\ee
where $Q$ is a parameter. For this to satisfy (\ref{odd}),
$f(\omega)$ must be purely imaginary.

In Sec.~\ref{sec:prelim}, we have defined a new field $\tF$, obtained
from $F$ by taking out some of the winding, cf. eq.~(\ref{tF}). 
The amount of winding taken out is
represented by $e^{iqx}$, where $q$ has so far been left undetermined. 
Regardless of the value of $q$, 
the condition (\ref{odd}) 
translates into the identical condition
for $\tF$ or, in terms of the real and imaginary parts of $\tF$,
into the condition that $R$ is odd, and $I$ is even about $x=0$. In other 
words,
\be
R(0,\omega) =  I'(0,\omega) = 0 \, ,
\label{bc0} 
\ee
where the prime denotes derivative with respect to $x$. 

As long as $F(x,\omega)$ is periodic, the condition (\ref{odd}) implies that
it is odd also under $P$ combined with a translation by the total length 
$L$, that is
\be
F(x,\omega) = - F^*(L-x,\omega) \, .
\label{odd_L}
\ee
We now choose $q$ so that  $\tF$, on the contrary, is even under this 
combination:
\be
\tF(x,\omega) = \tF^*(L-x,\omega) \, .
\label{even}
\ee
In terms of the real and imaginary parts of $\tF$, this becomes
\be
R'(L/2,\omega) = I(L/2,\omega) = 0 \, .
\label{bcJ}
\ee
For a periodic $F$, the restriction this imposes on $q$ is $e^{iqL} = -1$. We observe,
however, that we now have enough boundary conditions to formulate a boundary
problem on $[0,L/2]$ directly for $\tF$ (or, equivalently, for its real and imaginary
parts), so we can switch to more general twisted $F$ simply by abandoning the restriction,
i.e., by allowing for arbitrary values of $q$. In this way, $q$ becomes
a continuous parameter, 
which enters the calculation only through the expression (\ref{grad_tF}) for
the gradient term.

Next, we note that, for a given $q$, the boundary conditions (\ref{bc0}) and (\ref{bcJ})
still do not determine the solution uniquely. The reason is that the configuration space
is comprised of sectors corresponding to different winding numbers, and one can in
principle look for a solution to the boundary problem in each of these sectors 
individually. For example, for the uniform solutions (\ref{uni}),
\be
\tF(\ome,x) = f(\ome) e^{i (Q - q) x} \, ,
\label{uni_tF}
\ee
and the boundary conditions require only that $Q$ belongs to the discrete set
\be
Q = Q_m = q + \frac{\pi (2m + 1)}{L} \, ,
\label{Qm}
\ee 
where $m$ is any integer. The uniform states corresponding to different $Q_m$
in general have different energies, and
their activated decays are described by different critical droplets. We want to
make sure that we are solving for the right droplet, the one that describes the
decay of a particular uniform state. In other words, we want to be able to restrict to
one particular sector out of many.

Loosely speaking, the different sectors can be viewed as neighborhoods of different
uniform solutions, those with different $Q$ from the sequence (\ref{Qm}). 
More precisely, separation
into sectors can be done with the help of the winding number
\be
W(\omega) = \frac{1}{2\pi} \int_{-L/2}^{L/2} \partial_x \arg F(x,\omega) dx \, ,
\label{W}
\ee
which is well defined for any configuration for which the absolute value $|F(x,\omega)|$ 
does not vanish anywhere. As indicated in (\ref{W}), {\em a priori} 
one can imagine the winding number to be a function of $\omega$,
but here we restrict attention to solutions for which it is
the same for all $\omega$, $W(\omega) = W$.

For a uniform solution, $W = QL / 2\pi$. We can scan over all uniform solutions by
fixing $m$ in (\ref{Qm}) and varying $q$, and we now wish to concentrate on the
sector with $m=0$, for which
\be
Q = q + \frac{\pi}{L} \, .
\label{Q}
\ee
Recalling that by virtue of (\ref{bc0}) $f$ is purely imaginary,
we find the real and imaginary parts of (\ref{uni_tF}) to be
\ba
R(x,\ome) & = & - (\Im f) \sin \frac{\pi x}{L} \, , \label{uni_r} \\
I(x,\ome) & = & (\Im f) \cos \frac{\pi x}{L} \, . \label{uni_i}
\ea
These expressions reflect the characteristic property of the $m=0$ sector:
the real part of $\tF$ has no zeroes except the one 
at the origin, prescribed by (\ref{bc0}), and in addition,
at $x$ just above zero, the signs of the real and imaginary parts are opposite.
Indeed, for a given $q$, any field that has these properties 
and satisfies the boundary
conditions (\ref{bc0}) and (\ref{bcJ}) will have the same winding number as the uniform
solution (\ref{uni_r})--(\ref{uni_i}). This can be seen from the following relation,
which applies whenever $R$ has no zeroes at $x > 0$:
\[
2\pi W = q L + 2 \int_0^{L/2} \partial_x \arctan\frac{I(x,\ome)}{R(x,\ome)} dx 
\]
\be
{} = q L - \pi ~\sgn \frac{I(0, \ome)}{R(0^+ ,\ome)} \, .
\label{same_W}
\ee
Unlike the relative sign of $R$ and $I$, their overall sign is unimportant:
there is a companion set of solutions, with the same energy and current, 
that differ from the ones we find by an overall sign. 
As already noted after (\ref{odd}), this is a remnant of the original invariance 
with respect to global phase rotations. So, in what follows we identify the 
$m=0$ sector by the condition
\be
R(x,\ome)  >  0 \, , \hspace{3em} x > 0 \, . 
\label{Rpos}
\ee
Eq.~(\ref{same_W}) then shows that a solution (uniform or not) will be in the $m=0$ 
sector if $I(0, \ome) < 0$ and in the $m=-1$ sector if $I(0, \ome) > 0$. 

Let us note that the critical droplet responsible for fluctuations out of 
the $m=0$ sector can itself
be in either $m=0$ or $m=-1$ sector, depending on the sign of $q$. To see that,
let us first consider the special case $q=0$.
For the uniform solution, eq.~(\ref{Q}) now
gives $Q=\pi / L$, so the winding number is $W=1/2$. The uniform
solution with one fewer unit of winding (corresponding to $m=-1$) has
$Q= -\pi / L$ and $W=-1/2$. These solutions have exactly opposite currents and
equal energies. If we think of the twisted boundary conditions as being a result of
a magnetic flux through the ring, this special situation corresponds to the ring biased
by half the flux quantum. The critical droplet in this case has
$I(x,\omega) = 0$ identically, and the current on it is zero; 
because of the zero of $F$ at $x = 0$, the winding number is not well defined.
One can say that for $q=0$ the droplet lies at the boundary between the $m=0$ and
$m=-1$ sectors.

Recall that, in our computation, $q$ is a parameter that can be chosen arbitrarily.
As we change it from positive to negative, $I(0,\ome)$ changes from negative 
to positive, and the droplet moves from the $m=0$ sector to $m=-1$.
The underlying physics is that, at $q > 0$, the $m=0$
uniform solution is metastable, and the droplet describes its activated decay
to the $m=-1$ solution, which has a lower free energy. At $q=0$, the $m=0$ and
$m=-1$ uniform solutions become degenerate in energy,
and at $q<0$ the situation is reversed: for sufficiently small $|q|$, 
the $m=0$ solution is absolutely stable, and the droplet describes the decay of 
the $m=-1$ solution to it. Alternatively, even for $q < 0$, the droplet can
be interpreted as a fluctuation out of the $m=0$ sector. In that interpretation, its 
activation energy $\Fact$ will always be counted from the free energy of the $m=0$ 
uniform solution. This is how it has been presented on the chart of 
Fig.~\ref{fig:chart}, where the states corresponding to $q=0$ are shown
by the thick dashed line.

The sign condition (\ref{Rpos}), together with the boundary
conditions (\ref{bc0}) and (\ref{bcJ}), completes our formulation of the boundary problem
for $\tF$. Note that this boundary problem produces only half of the configuration,
namely, the part corresponding to $0 \leq x \leq L/2$. The other half can be obtained
by reflecting $R(x,\ome)$ as odd and $I(x,\ome)$ as even about $x=0$.

\section{Numerical results} \label{sec:num}
Before we proceed to the results, let us discuss some aspects of the substitution formula 
(\ref{eilen}) that are important for numerical work. Because $\lambda N(0)$ is 
constant, the cutoff integer $K$ in (\ref{eilen}) must increase with
decreasing temperature. On can express this by saying that, as $T$ is lowered, the
system builds up a ``synthetic'' frequency dimension, represented by a chain of
values of $k$. Physically, this dimension is associated with motion of electrons within
a pair. 

Suppose we take the smallest value of $K$, which we will
call $K_D$, to correspond to $T= T_c$. 
Then, from (\ref{eilen}),
\be
\frac{1}{\lambda N(0)} = \sum_{k=0}^{K_D - 1} \frac{1}{k + \half} \, .
\label{K_D}
\ee
Thus, we can use $K_D$ instead of $\lambda N(0)$ as a parameter determining 
the strength of the interaction. 
A typical value is $K_D = 10$, which corresponds to $\lambda N(0) = 0.234$. As long as 
$K_D$ is not much smaller than this, the results depend comparatively weakly on it.
As a consequence of (\ref{eilen}), once $K_D$ is specified, 
the size $K$ of the frequency dimension is related to $T/T_c$ as follows:
\be
\ln \frac{T}{T_c} = - \sum_{k = K_D}^{K - 1} \frac{1}{k + \half} \, ,
\label{TTc}
\ee
One limitation of this procedure is that, since
$K$ can only increase in discrete (integer) 
steps, $T/T_c$ is likewise limited to a discrete set of
values. The effect is significant mostly for $T$ near $T_c$. Indeed, 
the smallest $K > K_D$ one can
possibly choose is $K = K_D + 1$. Then, from (\ref{TTc}), the largest $T/T_c$ not equal to
unity is $T/T_c = \exp[-1/(K_D + \half)]$. For $K_D = 10$, this translates into the
cutoff at $T/T_c = 0.91$, seen in the chart of Fig.~\ref{fig:chart}.

Of course, one can choose $K=K_D$ to correspond not to $T=T_c$ but to a slightly lower
temperature. Since (\ref{K_D}) will no longer apply, this will result in a slightly 
different coupling, but will allow one to obtain  
a data point in the missing range of temperatures near $T_c$. Given
that the region near $T_c$ is not our main interest here, we will not pursue that.

\subsection{Units of length, energy, and current}
A natural unit of length in the present problem is the characteristic 
diffusion length $\bxi$ defined by
\be
\bxi^2 = \frac{\hbar D}{4\pi T_c} \, .
\label{bxi}
\ee
In this subsection only, we have restored $\hbar$.
For orientation, for $T_c = k_B \times (5~\mbox{K})$ and $D=1.2\times 10^{-4}$ m$^2/$s 
(values appropriate for amorphous MoGe wires \cite{Bezryadin:book}), $\bxi = 3.8$ nm.
In what follows, we will often quote the length $L$ in units of
$\bxi$, and the winding number parameter $q$ in units of $\bxi^{-1}$.

A natural unit of the free energy is 
\be
\fren_0 = (2\pi T_c)^2 N(0) A \bxi \, ,
\label{F_0}
\ee
where $A$ is the cross-sectional area of the sample. This amount represents a condensation
energy of order $2\pi T_c$ for each electron in a length $\bxi$ of the wire in 
an energy layer about $2\pi T_c$ thick near the Fermi surface. 
Similarly, a natural unit of the electric
current is 
\be
\cur_0 = 2e \times 2\pi T_c N(0) A D \bxi^{-1} = \frac{4e \fren_0}{\hbar} \, ,
\label{I_0}
\ee
where $e$ is the electron charge. 
In what follows, we present results for the free energy and 
the current in units of $\fren_0$ and $\cur_0$, respectively.

\subsection{Uniform solutions}
Uniform solutions are those of the form (\ref{uni}). For these, our main 
interest is in quantifying the accuracy of Bardeen's formula (\ref{bardeen}).
When we do not wish to refer to a specific length of the wire, $L$, we will use
$Q$ rather than $q$ as a parameter. If a value of $L$ is available, the two can 
be related by (\ref{Q}). As per discussion at the end of Sec.~\ref{sec:bc},
either is now considered a continuous parameter.

For a uniform solution, 
we do not need to solve the full
boundary problem: once $Q$ is chosen, the $x$-dependence of the solution is known,
and the Usadel equation becomes an equation for the amplitude $f(\ome)$. It reads
\be
\half D Q^2 f(\ome) + \frac{\ome f(\ome)}{[1 - |f(\ome)|^2]^{1/2}}
= 2\pi T \lambda N(0) \sum_{\ome > 0} f(\ome) \, ,
\label{usad_uni}
\ee
where $\ome$ takes the values
\be
\ome = 2\pi T (k + 1/2) \, , \hspace{3em} k = 0,\dots K - 1.
\label{ome}
\ee
Dividing (\ref{usad_uni}) by $2\pi T_c$, we see that the full set of parameters on which
the solution depends can be chosen as follows: $Q \bxi$, where $\bxi$ is the diffusion length 
(\ref{bxi}); $K_D$, which determines $\lambda N(0)$; and $K$, the upper cutoff in the
sum. From the latter two, the ratio $T/T_c$ can be calculated via (\ref{TTc}). In what
follows, we present the results as functions of that ratio, rather than $K$ itself.

We solve (\ref{usad_uni}) numerically by the multidimensional 
Newton-Raphson method. For given
$K_D$ and $T/T_c$, we compute the current as a function of the one remaining parameter,
$Q \bxi$. 
The current reaches a maximum at some ($T$-dependent) critical value $Q_c(T)$;
the value at the maximum is designated as the critical current, $I_c(T)$. The 
zero-temperature limit of $I_c(T)$ may be of independent interest, 
and we present results for it, for a few values of $K_D$ in Table~\ref{tab}. 
The table also lists the values of the coupling strength 
$\lambda N(0)$, computed from $K_D$ via (\ref{K_D}).
One may note the weakness of the dependence of $I_c$ on the coupling, especially 
for smaller
couplings.

\begin{table}
\begin{tabular}{|c|c|c|c|c|}
\hline
$K_D$ & 5 & 10 & 20 & 40 \\
\hline
$\lambda N(0)$ & 0.280 & 0.234 & 0.202 &  0.177 \\
\hline
$I_c(0) / \cur_0$ & 0.07752 & 0.07796 & 0.07819 & 0.07831 \\
\hline
\end{tabular}
\caption{The coupling strengths and the numerically obtained 
values of the critical current at $T=0$ for
several values of $K_D$. The current is in units of $\cur_0$, eq.~(\ref{I_0}). }
\label{tab}
\end{table}

\begin{figure}
\begin{center}
\includegraphics[width=3.4in]{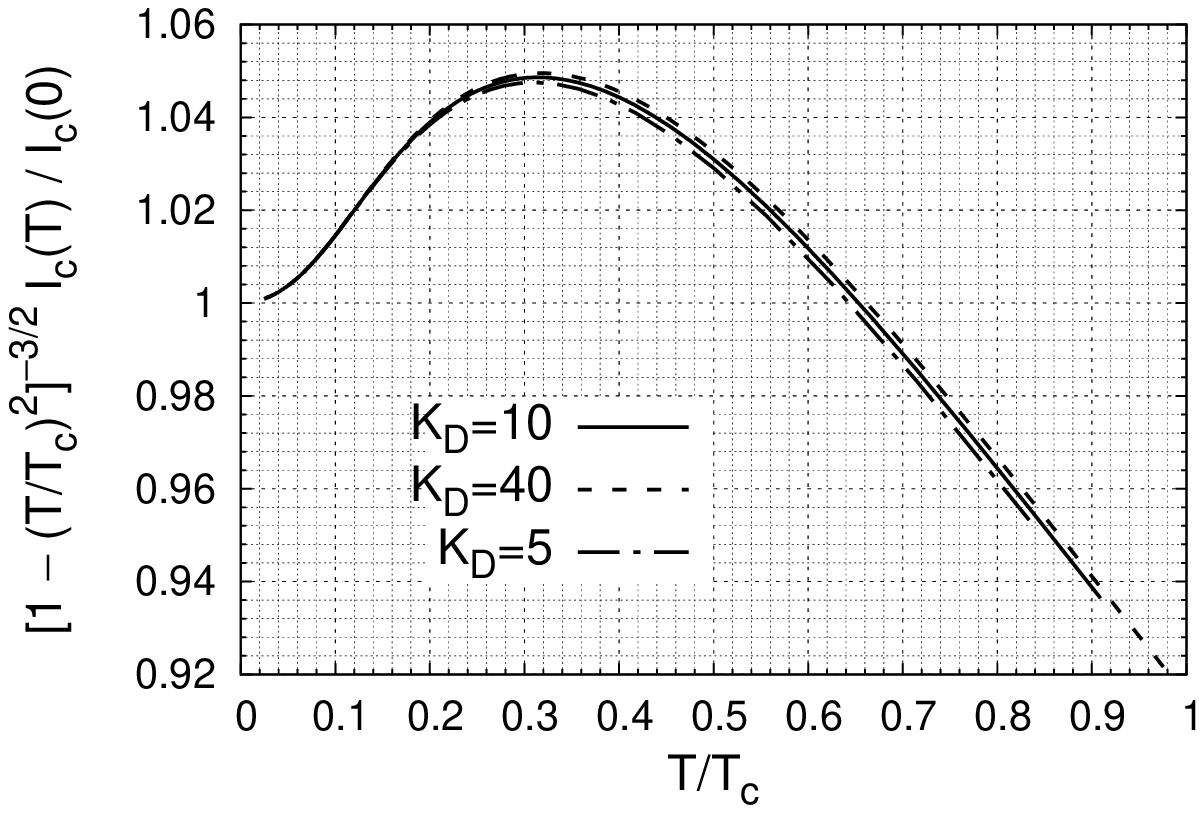}
\end{center}                                              
\caption{Correction factor to Bardeen's formula (\ref{bardeen}) for various coupling
strengths. The values of $K_D$ can be translated into coupling strengths by 
Eq.~(\ref{K_D}) or Table~\ref{tab}.
}                                              
\label{fig:corfac}                                                                       
\end{figure}

The ratio of the numerically computed $I_c(T)/I_c(0)$ to the interpolating function
$[1 - (T/T_c)^2]^{3/2}$ proposed by Bardeen \cite{Bardeen} 
is shown in Fig.~\ref{fig:corfac}. This ratio constitutes a correction factor 
to Bardeen's formula (\ref{bardeen}). The weak dependence of the result on $K_D$ means 
that, for weak coupling, the correction factor is essentially universal. 

One may observe that the correction factor retains a significant dependence on $T$ even
at the lowest temperatures: it grows by almost 2\% between $T=0$ and $T/T_c =0.1$.
This dependence, however, is almost entirely 
due to that in Bardeen's expression $[1 - (T/T_c)^2]^{3/2}$.
The numerically obtained $I_c(T)$ is
essentially flat for $T/T_c < 0.1$ and can be well approximated there by 
the corresponding value of $I_c(0)$ (as found in Table~\ref{tab}). 
This saturation of $I_c(T)$ at $T/T_c \approx 0.1$ is visible already on the chart
of Fig.~\ref{fig:chart} and is probably 
the main qualitative feature distinguishing the numerical result
from Bardeen's formula.

Next, we consider what happens to the uniform solution when $q$ is increased past
$q_c$, i.e., past the value at which the current reaches the maximum. 
As a point of comparison, we recall results for the theory of a single degree
of freedom (DOF), that is, a ``particle'' with a potential in the form of a cubic 
parabola, as used for instance in studies 
of Josephson junctions. The minimum of the potential corresponds to the ground state,
and the maximum to the critical droplet.
In that case, the maximum current is, at the same time, a bifurcation point at which
the droplet merges with the ground state. 

Turning to the present case, note that, for a uniform solution, the current
is proportional to the derivative of free energy density with respect to $q$,
$d \cF / dq$. Thus, $q = q_c$, at which the current is maximal, 
is an inflection point of $\cF$. For an infinite wire, the standard convexity argument
then guarantees that at $q = q_c$ the uniform solution becomes unstable
to spinodal decomposition. Finite-size effects can delay onset of the instability,
so for a wire of a finite length we expect that it occurs, if at all, 
at some $q = \qbif > q_c$. (Very recently, this effect has been observed experimentally
\cite{Petkovic&al}.)

\begin{figure}
\begin{center}
\includegraphics[width=3.4in]{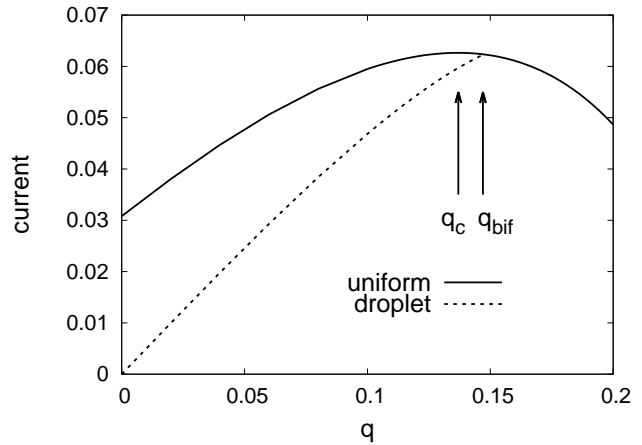}
\end{center}                                              
\caption{The current (in units of $\cur_0$) for the uniform solution (solid line) and
the critical droplet (dashed line), as a function of the winding number parameter $q$,
for $L/2 = 20.5$, $K_D = 10$, and $T/T_c = 0.4$. This plot illustrates the 
separation between the critical value $q_c$, where the current is
maximal, and the bifurcation point $\qbif$, where the solutions merge.
}                                              
\label{fig:merge}                                                                       
\end{figure}

As for a single DOF, the change in the stability properties (now at $q = \qbif$) is
accompanied by a bifurcation---a merging or splitting up of two or more solutions. 
In our case,
the solution merging with the uniform state is the critical droplet; the process
is illustrated in Fig.~\ref{fig:merge}. 
Note that, unlike for a singe DOF, the uniform solution continues to exist at
$q > \qbif$, even though it becomes absolutely unstable. As we discuss in 
Sec.~\ref{sec:top}, this can be attributed to the fact that there are two ``variants''
of the critical droplet, both merging with the uniform solution at $q = \qbif$. 
They are related by a discrete transformation which is a symmetry of the theory and 
so have the same free energy and current.

\subsection{Critical droplet}
The critical droplet is the saddle point of the free energy that sits at the top
of the potential barrier separating the uniform solution (\ref{uni}) with 
$Q = q + \pi/ L$, as given by (\ref{Q}),
from the uniform solution with one fewer unit of winding, i.e., $Q = q - \pi / L$.

The droplet is $x$-dependent, so finding it requires solving the full eq.~(\ref{usad}).
Numerically, we proceed by discretizing (\ref{usad}) on a one-dimensional spatial
grid of $N$ equally spaced sites (typically, $N=41$ or $N=81$) and applying the
multidimensional Newton-Raphson (NR) method to the resulting nonlinear
equations for $N\times K$ complex unknowns $F(x_j, \omega_k)$. As usual, the success
of the NR method depends on having a good initial approximation. We adopt the following
protocol. We start at a value of $K$ close to $K_D$, which corresponds to $T$ close to
$T_c$. There, we use the LA solution \cite{LA} of the GL theory as the initial 
approximation for the gap $\D(x)$, 
from which we reconstruct an approximation for $F(x,\ome)$ via
\be
F_{GL}(x,\ome_k) = \frac{\D(x)}{\ome_k} 
+ \frac{D \nabla^2 \D(x)}{2 \ome_k^2}  - \frac{|\D|^2 \D(x)}{2 \omega_k^3}
\label{FGL}
\ee
(the usual approximation for transitioning from the Usadel equation to the GL theory).
After the NR method finds the exact solution at this value of $K$, we compute the
exact $\D(x)$ and increase $K$ to
a larger value, using (\ref{FGL}) to populate the missing modes of $F$ (those with
$k$ between the old value of $K$ and the new one). That forms the initial approximation 
at the new $K$. In this way, increasing $K$ in steps, we move to progressively lower
values of the temperature.

Results of the computations have already been presented in Figs.~\ref{fig:chart} and
\ref{fig:merge}.

\subsection{Comparison to experiment} \label{subsec:exp}
Comparing our results to experiment requires, as the input data, the $T=0$ value of
the critical current $I_c(0)$, the critical temperature $T_c$, 
and the length $L$ of the sample. The weak dependence of
$I_c(0)$ on the coupling strength (see Table~\ref{tab}) allows us to determine
the unit $\cur_0$ of current by using the approximate relation $I_c(0) = 0.078 \cur_0$. 
The unit $\fren_0$ of free energy is then computed from (\ref{I_0}). Note that 
this does not require separate knowledge of the density of states and 
the cross-sectional area. With the value of $\fren_0$ in hand, we can refer to 
a chart such as that of Fig.~\ref{fig:chart} to find the activation barrier in physical
units.

\begin{figure}
\begin{center}
\includegraphics[width=3.4in]{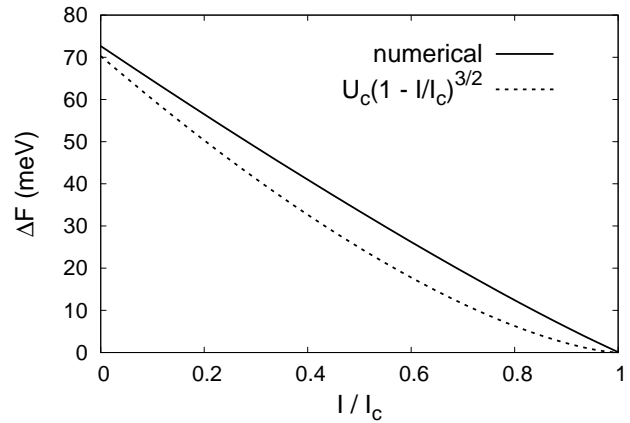}
\end{center}                                              
\caption{Solid line: 
the activation barrier $\Delta F$ for $T/T_c = 0.2$, 
$L=62$, $\lambda N(0) = 0.234$, and $\fren_0 = 159$ meV,
computed numerically. Dashed line: the function
$\Delta F = U_c (1 - I/I_c)^b$ with $U_c=70.3$ meV 
and $b = 3/2$, obtained in \cite{Aref&al} by fitting the experimental data for sample B.}
\label{fig:comp}                                                                       
\end{figure}

As an illustration, let us carry out this program for
sample B of ref.~\cite{Aref&al} (the longest wire described there). The parameters
as determined in \cite{Aref&al} are $I_c(0) = 12.11$ $\mu$A, $T_c = 5.48$ K, and $L = 221$ 
nm (for $T_c$, we use the value referred to in \cite{Aref&al} as $T_c'$).
Following the steps outlined above, we obtain $\fren_0 = 159$ meV. Next, we use numerical 
results for $L = 62$ (in units of $\bxi$), a somewhat smaller value than that
used for Fig.~\ref{fig:chart} but matching more closely the physical length. 
For illustration, we take $T/T_c = 0.2$.
The resulting activation barrier is plotted as a function of the equilibrium
current in Fig.~\ref{fig:comp}. The dashed line is the activation barrier obtained 
in \cite{Aref&al} by
fitting the experimental data. At small currents, the agreement is very good. Of main 
interest, however, is the region of large currents (say, those within 10\% of the critical), 
where the barrier becomes small enough for the switching transition to be observable.
There, the numerical result 
is almost a straight line and significantly overestimates the observed 
value. In fact, for $\D F = 6$ meV (the value obtained numerically for $I/I_c = 0.9$), 
the Boltzmann exponent at the temperature in question
is over 60; in all likelihood, such an exponent would render the transition unobservable.  

We can offer two possible explanations for the discrepancy at large currents.
One is that, the reasoning given in the Introduction notwithstanding,
nucleation of a critical droplet in a wire connected to superconducting leads does not
occur, even for large leads, exactly at fixed winding,
so the results obtained for a ring are not immediately applicable. 
The second possibility is that, 
at large currents, in addition to the uniform equilibrium states considered here,
there are other, possibly nonuniform, states available, activation from which proceeds
more easily.
For instance, for a periodic structure equivalent to a chain of Josephson junctions, the
$b=3/2$ scaling of $\Delta F$ at $I\to I_c$ applies even for rings of relatively
short lengths \cite{Kh:2016}. 
As the example of an array of phase-slip centers \cite{Skocpol&al}
indicates, such a nonuniform state may exist even in a morphologically uniform wire.

\section{Consequences of the translational invariance} \label{sec:top}
In addition to the symmetry under the discrete transformation (\ref{P}), 
the free energy is invariant, at an arbitrary fixed $q$, under the transformation
\be
P': \tF(x, \ome) \to -i \tF^*(L/2 - x, \ome) \, .
\label{P'}
\ee
This can be seen as a composition of (\ref{P}) with a translation 
by $L/2$ and a global phase rotation.
In terms of the real and imaginary parts of $\tF$, it corresponds to
\ba
R(x,\omega) & \to & - I(L/2 - x, \omega) \, , \label{tran_r} \\
I(x,\omega) & \to & - R(L/2 - x, \omega) \, , \label{tran_i}
\ea
The boundary conditions (\ref{bc0}) and (\ref{bcJ}) are also invariant under $P'$.
Indeed, this symmetry can be viewed as a discrete remnant of the full
transitional invariance that the theory had before we imposed 
(\ref{bc0}) and (\ref{bcJ}). As such, it is characteristic of the rather special 
situation presented by a uniform ring and is not available generically.

The uniform solution (\ref{uni_r})--(\ref{uni_i}) is invariant under $P'$, 
but the critical droplet is not. Thus, there must be a twin droplet, 
with the exact same values of the free
energy and current as the original, obtainable
from it by an application of $P'$. 
The only physical difference between the two
is that the ``core'' of the original droplet is at 
$x = 0$, while the core of the twin has been shifted to $x = L/2$.

One consequence of the existence of the twin is that, at the bifurcation point, there
are in fact not two but three
solutions merging. Thus, for instance, in Fig.~\ref{fig:merge}, the dashed line should now
be taken to mean two solutions: the original droplet and the twin. An index theorem can
then be used to relate the properties of the solutions before and after the
bifurcation. The argument is parallel to that used in \cite{Kuznetsov&Tinyakov}
to study bifurcation of a different type of solution, the periodic
instanton \cite{periodic}.

The requisite index theorem is the Morse equality \cite{Milnor}, which states that, 
for a smooth 
function $f$ with non-degenerate critical points on a compact smooth manifold $M$, 
\be
\sum_s N_s (-1)^s = \chi(M) \, ,
\label{index}
\ee
where the sum is over all the values of the index (the number of negative modes of $f$
at a critical point),
$N_s$ is the number of critical points of index $s$,
and $\chi(M)$ is the Euler characteristic of $M$. In our case, $M$ is
the product of $N\times K$ spheres of the form 
(\ref{nonlin}), and $f$ is the
discretized version of the free energy with the self-consistency condition
(\ref{scons}) and the expression (\ref{grad_tF}) substituted in. As a result, 
$f$ depends on $q$ as a parameter.

The critical points are, in our case, the various static solutions described
earlier. Assuming that no bifurcations except the one in question occur at 
$q = \qbif$, we can compute the change in the right-hand side of (\ref{index}) between
$q < \qbif$ and $q > \qbif$. That change must be zero,
as $\chi(M)$ is a topological invariant.
At $q < \qbif$, there are three solutions: the uniform solution of index 0, 
and the droplet with its twin, each of index 1. Their total contribution
to (\ref{index}) is $-1$. This is different from the case of a single DOF (the cubic
parabola), where only two solutions are merging, and the total index is zero.
In that case, the two solutions can simply disappear at $q=\qbif$.
In our case, the minimal structure needed at $q>\qbif$ to 
preserve the total index is a 
single solution with one negative mode. That is indeed what we have seen numerically:
the uniform solution remains but becomes absolutely unstable.

It is natural to ask where the instability that the uniform solution acquires at
$q > \qbif$ leads.
Numerically, this can be answered by displacing a little along the  
negative mode and following a relaxation algorithm. To better describe the results,
let us first recall that, in addition to the now unstable
uniform solution, which we will call $F_0$, there are, for the same $q$, uniform
solutions with other winding numbers. 
For the solution with one unit of winding fewer than $F_0$,
the real and imaginary parts of $\tF$ are
\ba
R(x, \omega) & = & g(\ome) \sin \frac{\pi x}{L} \, , \\
I(x, \omega) & = & g(\ome) \cos \frac{\pi x}{L} \, .
\ea
The difference with (\ref{uni_r})--(\ref{uni_i}) is that $R$ and $I$ now
have the same sign for $x > 0$. There are in fact two solutions of this form: 
one with $g > 0$, and the other with $g < 0$; let us call them $F_+$ and $F_-$.  
The transformation 
(\ref{tran_r})--(\ref{tran_i}) maps one into the other. 
Numerically, we have found that the instability of $F_0$
at $q > \qbif$ develops into $F_+$ or $F_-$, depending on the direction of
the initial displacement.\footnote{
It makes sense to consider two fields that differ only by an overall sign as
physically equivalent. Then, we would be talking about a loop, which starts at $F_+\sim F_-$, 
goes up to $F_0$, and then down back to $F_-\sim F_+$.}

Finally, we remark that, as $q$ is increased further, beyond $\qbif$, the uniform solution 
acquires additional negative modes and so is expected to go through additional
bifurcation points. We have not studied those in any detail.


\begin{thebibliography}{99}
\bibitem{Bardeen} J. Bardeen, Rev. Mod. Phys. {\bf 34}, 667 (1962).
\bibitem{Little} W. A Little, Phys. Rev. {\bf 156}, 396 (1967).
\bibitem{Sahu&al} M. Sahu, M.-H. Bae, A. Rogachev, D. Pekker, T.-C. Wei,
  N. Shah, P. M. Goldbart and A. Bezryadin, 
  Nature Physics {\bf 5}, 503 (2009).
\bibitem{Li&al} P. Li, P. M. Wu, Y. Bomze, I. V. Borzenets, G. Finkelstein, 
  and A. M. Chang, Phys. Rev. Lett. {\bf 107}, 137004 (2011).
\bibitem{Aref&al} T. Aref, A. Levchenko, V. Vakaryuk, and A. Bezryadin, 
  Phys. Rev. B {\bf 86}, 024507 (2012); 
  T. Aref, Ph.D. thesis, University of Illinois, 2010.
\bibitem{LA} J. S. Langer and V. Ambegaokar, Phys. Rev. {\bf 164}, 498 (1967).
\bibitem{Matveev&al} K. A. Matveev, A. I. Larkin, and L. I. Glazman,
  Phys. Rev. Lett. {\bf 89}, 096802 (2002).
\bibitem{Goswami&Chakravarty} P. Goswami and S. Charkavarty, 
  Phys. Rev. B {\bf 73}, 094516 (2006).
\bibitem{Kh:2016} S. Khlebnikov, Phys. Rev. B {\bf 94}, 064517 (2016)
  [arXiv:1604.07815].
\bibitem{Usadel} K. D. Usadel, Phys. Rev. Lett. {\bf 25}, 507 (1970).
\bibitem{Eilenberger} G. Eilenberger, Z. Phys. {\bf 214}, 195 (1968).
\bibitem{LO} A. I. Larkin and Yu. N. Ovchinnikov, Zh. Eksp. Teor. Fiz. {\bf 55},
  2262 (1968) [JETP {\bf 28}, 1200 (1969)].
\bibitem{Semenov&al}  A. V. Semenov, P. A. Krutitskii, and I. A. Devyatov,
  Pis'ma v ZhETF {\bf 92}, 842 (2010) [JETP Lett. {\bf 92}, 762 (2010)].
\bibitem{AL} L. G. Aslamazov and A. I. Larkin, Pis'ma v ZhETF {\bf 9}, 150 (1969)
   [JETP Lett. {\bf 9}, 87 (1969)].
\bibitem{KO} I. O. Kulik and A. N. Omelyanchuk, Pis'ma v ZhETF {\bf 21}, 216 (1975)
   [JETP Lett. {\bf 21}, 96 (1975)].
\bibitem{Marychev&Vodolazov} P. M. Marychev and D. Yu. Vodolazov, Pis'ma v ZhETF
  {\bf 103}, 458 (2016) [JETP Lett. {\bf 103}, 409 (2016)].
\bibitem{McCumber} D. E. McCumber, Phys. Rev. {\bf 172}, 427 (1968).
\bibitem{Petkovic&al} I. Petkovi\'{c}, A. Lollo, L.I. Glazman, and J.G.E. Harris, 
  Nat. Commun. {\bf 7}, 13551 (2016). 
\bibitem{Bezryadin:book}
  A. Bezryadin, {\em Superconductivity in Nanowires: Fabrication and Quantum Transport}
  (Wiley-VCH, Weinheim, 2013), Appendix A.
\bibitem{Skocpol&al} W. J. Skocpol, M. R. Beasley, and M. Tinkham, J. Low Temp. Phys. 
  {\bf 16}, 145 (1974).
\bibitem{Kuznetsov&Tinyakov}
  A. N. Kuznetsov and P. G. Tinyakov, Phys. Lett. B {\bf 406}, 76 (1997)
  [hep-ph/9704242].
\bibitem{periodic} S. Y. Khlebnikov, V. A. Rubakov, and P. G. Tinyakov, 
  Nucl. Phys. B {\bf 367}, 334 (1991).
\bibitem{Milnor} J. Milnor, {\em Morse Theory}
  (Princeton University Press, Princeton, 1963).
\end{thebibliography}
\end{document}